\documentclass{aastex}
\usepackage{spr-astr-addons}

\newcommand{\vth}{\ensuremath{v_{\text{th}}}}
\newcommand{\kB}{\ensuremath{k\!_B}}
\newcommand{\eps}{\ensuremath{\varepsilon}}

\newcommand{\be}{\begin{equation}}
\newcommand{\ee}{\end{equation}}
\newcommand{\bs}{\begin{subequations}}
\newcommand{\es}{\end{subequations}}

\newcommand{\pa}{\ensuremath{_\parallel}}
\newcommand{\se}{\ensuremath{_\perp}}

\newcommand{\uint}{\ensuremath{\int_{-\infty}^\infty}}

\newcommand{\f}[1]{\ensuremath{\boldsymbol{#1}}}

\newcommand{\pd}[2][]{\ensuremath{\frac{\partial #1}{\partial #2}}}

\newcommand{\df}{\ensuremath{\mathrm{d}}}

\begin{document}
\title{A New Distribution Function for Relativistic Counterstreaming Plasmas}
\shorttitle{Relativistic Counterstreams}
\shortauthors{Tautz}

\author{R.\,C. Tautz\altaffilmark{1}} 
\email{R.C.Tautz@uu.nl}

\altaffiltext{1}{Astronomical Institute Utrecht, Princetonplein 5,\\
NL-3584CC Utrecht, Netherlands}

\begin{abstract}
The particle distribution function that describes two interpenetrating plasma streams is re-investigated. It is shown how, based on the Maxwell-Boltzmann-J\"uttner distribution function that has been derived almost a century ago, a counterstreaming distribution function can be derived that uses velocity space. Such is necessary for various analytical calculations and numerical simulations that are reliant on velocity coordinates rather than momentum space. The application to the electrostatic two-stream instability illustrates the differences caused by the use of the relativistic distribution function.
\end{abstract}


\section{Introduction}

Plasma physics---both MHD (magnetohydrodynamics) and kinetic theory---are based on the knowledge of a distribution function that provides statistics about the average velocity direction (bulk flow) and the deviation from that mean value (known as temperature). Whereas MHD calculations are based on the use of a Maxwellian velocity distribution, such is not the case in kinetic theory, where the distribution function can be calculated using, e.\,g., the Vlasov equation. Kinetic theory \citep[e.\,g.,][]{rs:rays} is mostly used when dilute plasmas are considered that do not satisfy the condition of frequent binary particle collisions so that no Maxwellian velocity distribution is established.

In various and already historical work, the classic Maxwellian velocity distribution has been generalized to a relativistic gas (see, e.\,g., \citealt{jut11:dyn}; \citealt{jut11:max}; \citealt{syn57:gas}; \citealt{fra79:col}). The result was the MBJ (Maxwell-Boltzmann-J\"uttner) distribution function (\citealt{jut11:dyn}; \citealt{jut11:max}), which is essentially given through
\be
f\propto\exp\!\left(-\alpha\sqrt{1+\f p^2}\right)
\ee
where $\alpha$ is a temperature-related parameter.

However, there are cases when the use of a distribution function in momentum space is (i) not suitable for the analytical calculations at hand; and/or (ii) not implemented in the numerical (simulation) code. A recent example is the investigation of LIDAR (LIght Detection And Ranging) Thomson scattering systems in ITER (International Thermonuclear Experimental Reactor) plasmas \citep{bea08:tho}. The calculation was based on a formula \citep{hut87:pla} for the scattered power per unit solid per unit angular frequency, and thus the particle distribution function had to be expressed in terms of velocity variables rather than momentum variables.

Another example is the PIC (Particle-In-Cell) simulation code \textsc{Tristan} (originally \citealt{bun93:pic}, see also, e.\,g., \citealt{sak04:mag}), which is used for the investigation of plasma instabilities in the context of astrophysical scenarios such as the generation of magnetic fields at shock wave sites. A basic example of such instabilities is the generalized filamentation (or, originally, Weibel) instability (\citealt{wei59:wei}; \citealt{fri59:wei}).

In this short Note, particle distribution functions are considered that describe two interpenetrating plasma streams. Such distributions are widely used in plasma astrophysics, because in the right reference frame, all outflow motion into ambient media and shock wave sites can be described by counterstreaming flows \citep[e.\,g.,][]{tau05:cov}. Examples are solar, stellar, and galactic winds, and the interaction of relativistic jets such as that from GRBs (gamma-ray bursts) and AGNs (active galactic nuclei) with the interstellar medium. Especially in the latter cases, one needs a distribution function that accounts for the relativistic effects, which can modify the instability rates significantly \citep{usr08:wei}.

In Sec.~\ref{dist}, it will be shown how the relativistic distribution function for counterstreams in momentum space, which has been derived \citep{tau05:cov} from the MBJ distribution above, can be rewritten in terms of velocity variables. In Sec.~\ref{twostr}, the application to the two-stream instability will illustrate the significant differences between the relativistic and the non-relativistic counterstreaming distribution. Finally, the results are summarized in Sec.~\ref{summ}.

\section{Counterstreaming Distribution}\label{dist}

\begin{figure*}
\includegraphics[width=125mm]{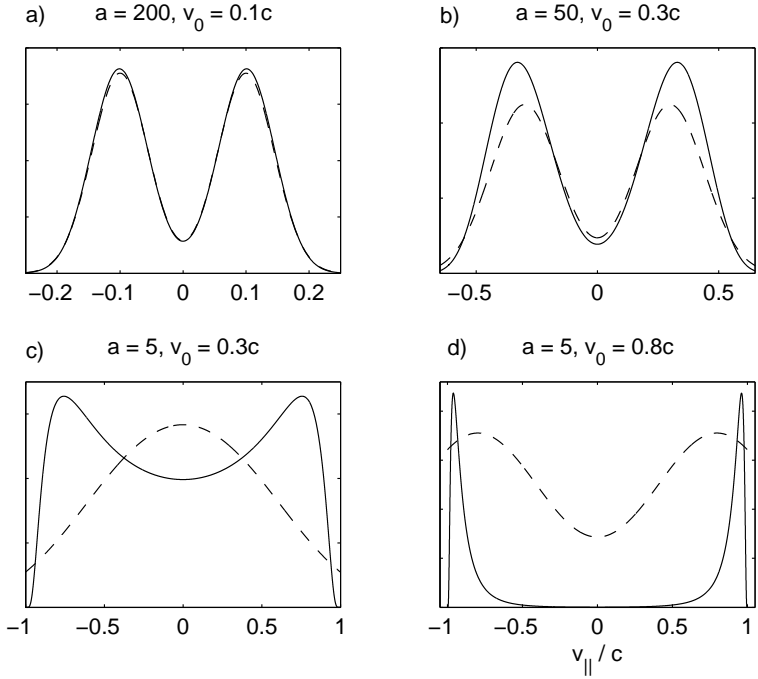}
\caption{The counterstreaming distribution function as a function of $v\pa/c$ for different values of the normalized temperature $\alpha$ and the streaming velocity $v_0$. Shown are the relativistic version (solid line) from Eq.~\eqref{eq:dist} in comparison to the non-relativistic version (dashed line) from Eq.~\eqref{eq:ndist}. Note that, for small temperatures $\kB T\ll mc^2$ and small streaming velocities $v_0\ll c$, the two distributions agree (panel~a), whereas, even for $\kB T\lesssim mc^2$, no agreement at all can be seen (panels~c and~d). Note also that, in panel~d, the non-relativistic distribution is superelevated by a factor of $50$.}
\label{ab:dist}
\end{figure*}

In momentum space, the two components of the relativistic counterstreaming distribution function can be constructed intuitively from the Maxwell-Boltzmann-J\"uttner distribution function (\citealt{jut11:dyn}; \citealt{jut11:max}; \citealt{sch04:cov};\\
\citealt{usr05:cov};\\
\citealt{tau05:cov}) as
\be\label{eq:dist}
f^\mp=C\exp\Bigl[-\alpha\sqrt{1+(p\pa\mp p_0)^2+p\se^2}\,\Bigr],
\ee
where $\alpha=mc^2/(\kB T)$ is proportional to the inverse temperature. By integrating over the whole velocity space, the normalization constant $C$ can be evaluated \citep{sch04:cov} to
\be\label{eq:C}
C=\frac{\alpha}{4\pi(mc)^3K_2(\alpha)},
\ee
where $K_2$ denotes the modified Bessel function of the second kind of order two \citep[a.\,g.,][]{ab:math}. Note that Eq.~\eqref{eq:dist} does not allow for anisotropic temperatures (i.\,e., thermal velocities) in the directions parallel and perpendicular to the counterstream. There are several ways to implement such temperature anistropies in covariant (see, e.\,g., \citealt{yoo89:wei};\\
\citealt{sch04:cov}; \citealt{tau05:cov}) and semi-relativistic (e.\,g., \citealt{zah07:sem};\\
\citealt{tau08:wei}) distribution functions; however, there is no general agreement as to which form is best suited. However, temperature anisotropies become increasingly negligible for significant counterstreaming velocities \citep{tau07:spo}. Therefore, it is appropriate especially for relativistic distributions, to neglect temperature anistropies at least for the present.

To transform the distribution function from Eq.~\eqref{eq:dist} to velocity space, use ise made of the fact\\
\citep{fra79:col} that
\be
p^2\df^3p=\gamma^5v^2\,\df^3v,
\ee
where $p$ and $v$ are momentum and velocity components, respectively, and where
\be
\gamma=\frac{1}{\sqrt{1-\left.\left(v\pa^2+v\se^2\right)\right/c^2}}
\ee
is the relativistic Lorentz factor. Thus, the transformed distribution function in velocity space reads
\begin{eqnarray}
&&f^\mp=C\,\gamma^5\exp\Biggl\{-\alpha\Biggl/\nonumber\\
&&\hspace{-1.5em}\sqrt{1-\left(c^2-v\pa v_0\right)^{-2}\left[\left(v\pa\mp v_0\right)^2+v\se^2\left(1-v_0^2\right)\right]}\Biggr\}.
\end{eqnarray}
where $v\pa,v\se,v_0\in\left[-c,c\right]$ and where the total velocity,
\be
v_{\text{all}}=\left(c^2\mp v\pa v_0\right)^{-2}\left[\left(v\pa\mp v_0\right)^2-v\se^2\left(1-v_0^2\right)\right]
\ee
is also limited by the speed of light, $c$. Note that the correct relavistic addition theorem for velocities has been used and that the perpendicular velocity is decreased if the counterstreaming velocity, $v_0$, is increased.

For small velocities and for non-relativistic temperatures, i.\,e., $v_0\ll c$ and $\vth\ll c$, a series expansion of both the argument of the exponential function and the Bessel function yields the well-known non-relativistic version of the distribution function \citep{tau05:cov}, namely
\be\label{eq:ndist}
f^\mp_{\text{nr}}=\left(\vth\sqrt\pi\right)^{-3}\exp\!\left[-\frac{\left(v\pa\mp v_0\right)^2+v\se^2}{\vth^2}\right],
\ee
where $\vth=\sqrt{2\kB T/m}=\sqrt{2/\alpha}$ is the non-relativistic thermal speed, which is proportional to the inverse square root of the temperature parameter $\alpha$ used for the relativistic distribution.

\begin{figure}[t]
\includegraphics[width=82mm]{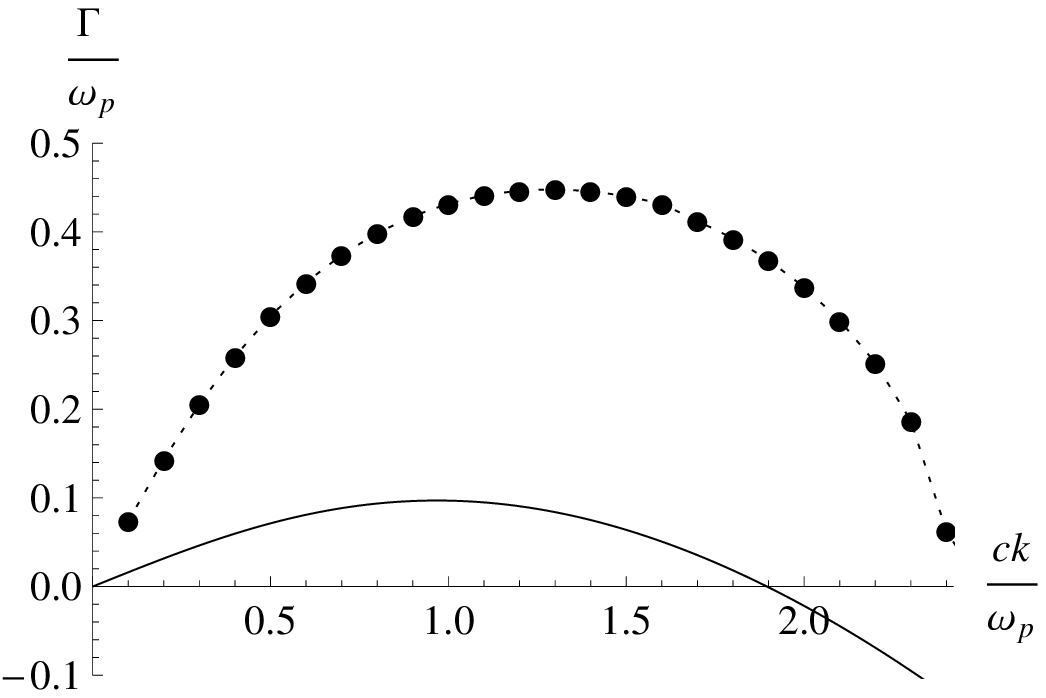}
\caption{The growth rate for the longitudinal mode of the two-stream instability, which is the solution of Eq.~\eqref{eq:twostr}. The solid line shows the case of the non-relativistic function from Eq.~\eqref{eq:ndist}, whereas the dots show the relativistic version, which uses Eq.~\eqref{eq:dist}. The solution of the dispersion relation for the relativistic distribution function is, numerically, both more delicate and time-consuming and has therefore been limited to a few sample points, indicated by the dots.}
\label{ab:growth}
\end{figure}

The total distribution function that describes two counterstreaming plasma components is then combined as
\be\label{eq:rdist}
f=\eps f^-+\left(1-\eps\right)f^+,
\ee
where $\eps\in\left[0,1\right]$ denotes the relative intensity of forward and backward stream. Both the parameters $\alpha$ and the counterstream velocity, $v_0$, can be allowed to have different values for the forward and the backward stream. In that case, however, $v_0$ becomes dependent on $\alpha$ (or vice verse), because of the overarching condition of vanishing zero-order current that has to be fulfilled \citep{tau07:wea}.

Furthermore, the integral over the velocity space is then transformed as
\be
\int\df^3v=2\pi\int_{-c}^c\df v\pa\int_0^{\sqrt{c^2-v\pa^2}}\df v\se\,v\se
\ee
so that the distribution function $f$ from Eq.~\eqref{eq:rdist} is normalized to unity. In that context, note that the normalization factor $C$ from Eq.~\eqref{eq:C}, that was derived for the distribution function in momentum space, Eq.~\eqref{eq:dist}, remains unchanged throughout the transformation to velocity space.

\section{Two-Stream Instability}\label{twostr}

To illustrate the use of the relativistic counterstreaming distribution function, Eq.~\eqref{eq:dist}, the electrostatic two-stream instability (see, e.\,g., \citealt{sch04:cov}; \citealt{tau05:co1}) is evaluated both for the non-relativistic and the relativistic distribution functions. Parameters are chosen as $v_0=0.5\,c$ and $\alpha=10$, thus one has $\vth=0.45\,c$. The relative intensity of the two plasma streams is set to be equal to $1/2$. For the longitudinal mode, the dispersion relation is given through
\be\label{eq:twostr}
\omega=\omega_p^2\uint\df p\pa\int_0^\infty\df p\se\;\frac{p\se p\pa}{\gamma\left(k\pa v\pa-\omega\right)}\,\pd[f]{p\pa},
\ee
where $\omega_p=\sqrt{4\pi nq^2/m}$ is the plasma frequency.

In the case of the non-relativistic distribution function, the double integral in Eq.~\eqref{eq:twostr} can be expressed in terms of the plasma dispersion function or $Z$ function of \citet{fri61:pla} \citep[see also][]{tau05:co1}, whereas such is hardly possible for the relativistic distribution function. Here, numerical integration methods are used instead.

The numerical solution of the dispersion relation is shown in Fig.~\eqref{ab:growth}. Obviously, the growth rate, which is defined as the (positive) imaginary part of the frequency $\omega(k)$, exceeds is significantly increased if the relativistic distribution function is used. Furthermore, both the wavenumber of the maximum growth rate, $k_{\text{max}}$, and the wavenumber marking the end of the unstable range, $k_{\text{end}}$, are both larger than for the non-relativistic distribution. This can be understood by keeping in mind that, from Fig.~\ref{ab:dist}, one knows that the anisotropy is much more clear-cut for the relativistic distribution.

\section{Summary and Conclusion}\label{summ}

It has been known for a long time that, for relativistic temperatures (or, to be more accurate, for relativistic thermal velocities), the classic Maxwellian distribution function becomes increasingly inaccurate. Such is escpecially the case for distribution functions that describe plasmas streaming with high velocities, as is the case for two counterpropagating components. Because there exist wide application ranges of such distributions, it is both appropriate and necessary to take care of a precise basic construction.

In this short Note, it has been shown how a particle distribution function that describes two interpenetrating plasma streams can be generalized to relativistic velocities. Based on the MJB distribution function, an expression has been derived that explicitely uses velocity coordinates. Such is advantageous both for analytical calculations and for numerical simulations that rely on velocity coordinates.

As an example to illustrate the application of the relativistic counterstreaming distribution and, at the same time, to demonstrate the both qualitative and quantitative differences that result from the use of a correct relativistic distribution function, the electrostatic two-stream instability has been chosen. It has been shown that, due to the sharply emphasized anisotropy profile of the relativistic distribution function, the growth rate is significantly higher and the unstable wavenumber range is more extended as was the case for the non-relativistic distribution.

In future work, the distribution function proposed here should be applied mainly to numerical simulations. The differences between non-relativistic, semi-relativistic, and fully relativistic initial distribution functions have to be worked out. Only then can one assess the necessity of a, for analytical calculations rather unwieldy, distribution function as that proposed here.

\acknowledgments
This work was supported by the Deutsche Forschungsgemeinschaft (DFG) through grant No.\ Schl~201/19-1 and by the German Academy of Natural Scientists Leopoldina Fellowship Programme through grant LDPS 2009-14 funded by the Federal Ministry of Education and Research (BMBF).


\end{document}